\documentclass[RNAAS]{aastex62}

\begin{document}

\title{\textit{Gaia} EDR3 confirms a red dwarf companion of the nearby F1 star HD 105452 
and reveals a new brown dwarf companion of the M4.5 dwarf SCR J1214-2345}

%% Note that the corresponding author command and emails has to come
%% before everything else. Also place all the emails in the \email
%% command instead of using multiple \email calls.
\correspondingauthor{Ralf-Dieter Scholz}
\email{rdscholz@aip.de}

\author[0000-0002-0894-9187]{Ralf-Dieter Scholz}
\affiliation{Leibniz Institute for Astrophysics Potsdam (AIP),\\
An der Sternwarte 16, 14482 Potsdam, Germany}

\begin{abstract}
There are 88 stars which lack colours, but have measured parallaxes in 
\textit{Gaia} EDR3 that place them within 20\,pc from the sun. Among them 
we found two new common parallax and proper motion (CPPM) companions separated 
from their primaries by about 3\,arcsec. The CPPM companion of a nearby 
($d=14.98$\,pc) F1 star, \object{HD 105452 B}, was already imaged 
with the \textit{Hubble Space Telescope} and is now confirmed with \textit{Gaia} 
data and photometrically classified by us as M4 dwarf.
The other CPPM companion, \object{SCR J1214-2345 B} orbiting an M4.5
dwarf at $d=10.77$\,pc, represents the faintest brown dwarf discovery 
made by \textit{Gaia} so far. It was also imaged by the VISTA Hemisphere Survey
and partly detected in the near-infrared. Our photometric 
classification led to an uncertain spectal type of T1$\pm$3 and
needs to be confirmed by spectroscopic follow-up.
\end{abstract}

%% Note that RNAAS manuscripts DO NOT have abstracts.

%% Start the main body of the article. If no sections in the 
%% research note leave the \section call blank to make the title.
\section{Investigated \textit{Gaia} EDR3 20\,pc sub-sample} 

The \textit{Gaia} Early Data Release 3 \citep[EDR3;][]{2020arXiv201201533G} provided
parallax measurements for about 1.5 billion stars. Selecting only the 
nearest stars with the largest parallaxes taking into account their errors, 
we found 2756 stars with $(Plx+3\times{e\_Plx})>50$\,mas comprising the EDR3
20\,pc sample. There are 88 stars without $G-RP$ colour measurements among
those 2756 stars. The majority (48) of them are very faint ($19.2<Gmag<20.9$)
objects in the Galactic plane ($|b|<9\deg$), where \textit{Gaia} measurements are
problematic. Five objects were according to SIMBAD brown dwarfs with spectral
types between L4.5 and L6.5, whereas the remaining 35 represented the
fainter components of close common parallax and proper motion (CPPM) pairs
in \textit{Gaia} EDR3. These CPPM pairs with projected separations between 0.7\,arcsec
and 3.5\,arcsec were mostly already known in the Washington Double Star (WDS) 
catalogue \citep{2001AJ....122.3466M}. They typically contained M dwarf secondaries, 
but also included an M3+L0 and an L1.5+L4.5 pair. With such small separations, 
the \textit{Gaia} EDR3 proper motions of these nearby CPPM pairs disagreed to some 
extent because of the effect of orbital motion. Only three of the 35 CPPM
pairs had no WDS entry. One of them contains the known DQ6 white dwarf 
(WD) \object{GJ 86 B} that did not have a \textit{Gaia} DR2 parallax. Hence it
was not yet included in the \textit{Gaia} DR2 20\,pc WD sample of 139 objects
of \citet[][see their Table 4]{2018MNRAS.480.3942H}. 
The other two are described below.

\section{Confirmed companion of a nearby F1 star}

A relatively bright star \object{Gaia EDR3 3489338019474046720},
hereafter \object{HD 105452 B}
($Gmag\approx11.70, Plx=66.23\pm0.17$\,mas, 
$pmRA=+23.99\pm0.29$\,mas/yr, $pmDE=-59.66\pm0.45$\,mas/yr),
was measured at an angular separation of 3.1\,arcsec from
the known very bright F1 star \citep{2006AJ....132..161G} 
\object{HD 105452} = \object{Gaia EDR3 3489338019475637760}
($Gmag\approx3.95, Plx=66.77\pm0.18$\,mas,
$pmRA=+96.98\pm0.18$\,mas/yr, $pmDE=-40.02\pm0.21$\,mas/yr).
Whereas the parallaxes almost agree within their errors,
the relatively small proper motions are rather different. The
total proper motion difference is about three times larger than the
expected effect of orbital motion ($\approx$25\,mas/yr) if we
assume a system mass of two solar masses and a circular orbit in
the plane of the sky \citep[cf.][]{2018A&A...613A..26S}. This may
indicate that our simple assumptions are not correct or that the
system is already dissolving. Because of the missing colour measurements
in \textit{Gaia} EDR3, \object{HD 105452 B}, with its absolute magnitude of 
$M_G=10.80$\,mag, could be a hot WD similar to \object{Sirius B} 
or a red dwarf.

Fortunately, \object{HD 105452} was selected by \citet{2014ApJ...784..148D} 
as a point spread function (PSF) reference star in their investigation of
a debris disk around another star. They mentioned an apparent companion
at a separation of 2.2\,arcsec, when they observed \object{HD 105452}
with the \textit{Hubble Space Telescope (HST)} advanced camera for surveys (ACS)
using the high resolution channel (HRC). Figure \ref{fig:1}a shows that
this companion, marked as \object{HD 105452 B}, appeared relatively bright 
in the observation with the red F814W filter. \citet{2014ApJ...784..148D}
noted problems caused by this obviously red companion in their applied PSF 
subtraction technique with the F814W filter, whereas these problems 
were less acute with the F435W and F606W filters. We conclude that 
\object{HD 105452 B} is not a WD but a red dwarf.
According to \citet[][their Table 7]{2020A&A...642A.115C}, the average
absolute magnitude of M4 dwarfs in \textit{Gaia} DR2 was $M_G=10.88$\,mag. Therefore, 
we assign a photometric spectral type of M4V to \object{HD 105452 B}. 

\section{New brown dwarf companion}

The faint CPPM object \object{Gaia EDR3 3489874340630661248}, hereafter
\object{SCR J1214-2345 B} 
($Gmag\approx20.58, Plx=94.18\pm1.08$\,mas,
$pmRA=+58.67\pm1.17$\,mas/yr, $pmDE=+53.39\pm0.94$\,mas/yr),
was measured next (separation 3.5\,arcsec) to the 
nearby M4.5 dwarf \citep{2006AJ....132..866R} 
\object{SCR J1214-2345} = \object{Gaia EDR3 3489874340631095936}
($Gmag\approx12.29, Plx=92.89\pm0.03$\,mas,
$pmRA=+44.18\pm0.03$\,mas/yr, $pmDE=+84.32\pm0.04$\,mas/yr).
The parallaxes of this CPPM pair are in good agreement, and the total 
proper motion difference is not so large as in case of \object{HD 105452 AB} -
only about 60\% larger than the expected orbital motion effect of 
about 20\,mas/yr for an assumed system mass of only 0.5 solar 
masses. An $RP$ magnitude was not given in \textit{Gaia} EDR3, whereas the 
$BP\approx20.4$\,mag should be taken with caution as it was based
on only two measurements and probably affected by the close
primary.

\begin{figure}[h!]
\begin{center}
\includegraphics[scale=0.56,angle=0]{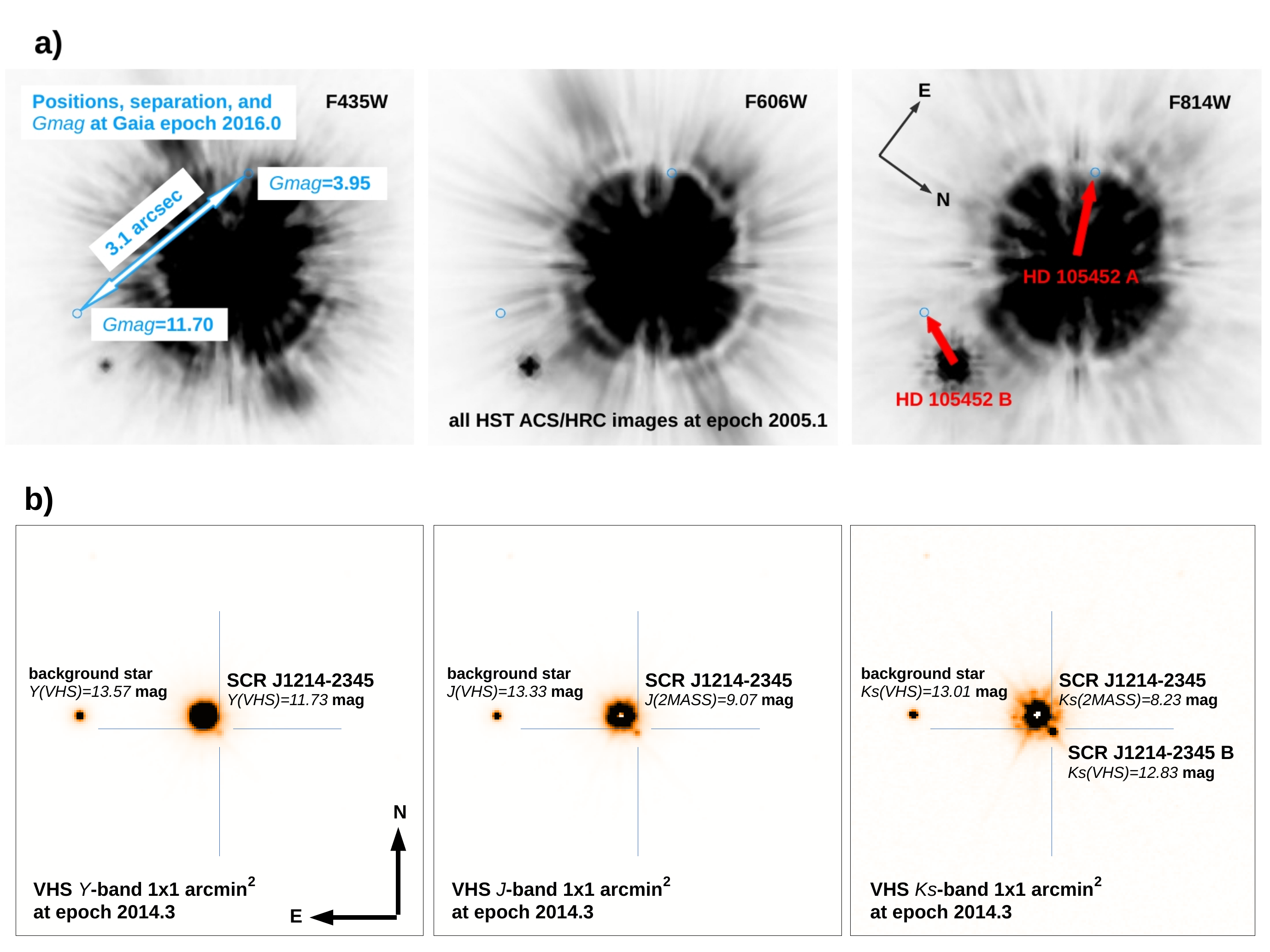}
	\caption{\textbf{a)} \textit{HST} ACS/HRC images of \object{HD 105452}, used 
	as PSF star in \citet{2014ApJ...784..148D}, from the
	original observing program 10244 of Mark Wyatt (PI) extracted 
	from \url{https://archive.stsci.edu/hst/preview/}.
	The three exposures with different filters (left: F435W, middle: 
	F606W, right: F814W) were obtained with a coronographic mask 
	of 1.8\,arcsec. The positions of two objects detected in \textit{Gaia} EDR3 
	are overplotted with blue open circles. In the left panel, 
	their separation and $G$ magnitudes are overlayed. The red arrows in 
	the right panel illustrate the proper motions of 
	the two components over the time baseline of about 11 years, which are 
	consistent with the \textit{Gaia} EDR3 measurements and probably indicate 
	orbital motion. 
	\textbf{b)} from left to right: VHS $Y$-, $J$-, and $K_s$-band images
	(epoch 2014.3) centered on the \textit{Gaia} EDR3 position (epoch 2016.0)
	of \object{SCR J1214-2345 B}.
	The VHS catalogue position of \object{SCR J1214-2345 B} is only 137\,mas
	away from its \textit{Gaia} EDR3 position and consistent with its proper motion.
	The measured VHS (AperMag3) magnitudes of \object{SCR J1214-2345}, 
	\object{SCR J1214-2345 B}, and 
	a background star are marked. However, 2MASS $J$ and $K_s$ magnitudes of (the
	unresolved) \object{SCR J1214-2345} are given, because its VHS catalogue
	magnitudes appeared much to faint. \label{fig:1}}
\end{center}
\end{figure}

Optical and near-infrared images of \object{SCR J1214-2345} can be 
found in the VISTA Hemisphere Survey \cite[VHS;][]{2013Msngr.154...35M}. The
extracted 1$\times$1\,arcmin$^2$ $YJK_s$ images centered on the \textit{Gaia} EDR3 position
of \object{SCR J1214-2345 B} are shown in Figure \ref{fig:1}b. The small
proper motion can hardly be seen because of the epoch difference of less
than two years. The VHS catalogue lists only $K_s=12.83$\,mag (AperMag3),
but no $YJ$ magnitude measurements for \object{SCR J1214-2345 B}, although
one can see its fainter counterparts in the images, where we also marked a
background star with simlar $K_s$ but brighter $YJ$ magnitudes for comparison.
Using the relations between absolute magnitudes and spectral types given
in \citet{2018A&A...619L...8R} led to a spectral type of 
T4-T5 from $M_G=20.45$\,mag but only L8 from $M_{K_s}=12.70$\,mag. 
Therefore, we can only provide a very uncertain photometric classification 
of \object{SCR J1214-2345 B} as T1$\pm$3 dwarf. However, it is an important
addition to the 20\,pc census of LTY dwarfs of \citet{2020arXiv201111616K}
and the faintest new brown dwarf discovery from 
using \textit{Gaia} data so far \citep[cf.][]{2020A&A...637A..45S}.
 
%% An example figure call using \includegraphics

\acknowledgments

This work presents results from the European Space Agency (ESA) space mission \textit{Gaia}. 
\textit{Gaia} data are being processed by the \textit{Gaia} Data Processing and Analysis Consortium 
(DPAC). Funding for the DPAC is provided by national institutions, in particular 
the institutions participating in the \textit{Gaia} MultiLateral Agreement (MLA).
The \textit{Gaia} mission website is \url{https://www.cosmos.esa.int/gaia}.
The \textit{Gaia} archive website is \url{https://archives.esac.esa.int/gaia}.

\end{document}